\documentclass{pasa}

\title[SEDs of Systems with Forming Companions]{The Influences of Forming Companions on the Spectral Energy Distributions of Stars with Circumstellar Discs}

\author[O.V. Zakhozhay]{Olga V. Zakhozhay \thanks{E-mail: zakhozhay.olga@gmail.com}\\
\affil{Main Astronomical Observatory, National Academy of Sciences of Ukraine, Kyiv, 03143, Ukraine}
}

\jid{PASA}
\doi{10.1017/pas.\the\year.xxx}
\jyear{\the\year}

\usepackage[authoryear]{natbib}
\bibpunct{(}{)}{;}{a}{}{,}
\begin{document}

\begin{abstract}
We study a possibility to detect signatures of brown dwarf companions in a circumstellar disc based on spectral energy distributions (SED). We present the results of SED simulations for a system with a 0.8~$M_{\odot}$ central object and a companion with a mass of 30~$M_J$ embedded in a typical protoplanetary disc. We use a solution to the one-dimensional radiative transfer equation to calculate the protoplanetary disc flux density and assume, that the companion moves along a circular orbit and clears a gap. The width of the gap is assumed to be the diameter of the brown dwarf Hill sphere. Our modelling shows that the presence of such a gap can initiate an additional minimum in the SED profile of a protoplanetary disc at $\lambda =  10 - 100$~$\mu$m. We show that the depth of this minimum and the wavelength of the maximum difference between the SEDs of the system with and without a companion are related to the companion mass and its proximity to the star. We found that it is possible to detect signatures of the companion when it is located within 10~AU, even when it is as small as  3~$M_J$.  We also analyse how the disc parameters (the inner radius and the temperature profile) change the maximum difference between the SEDs for the same systems with and without a companion. The SED of a protostellar disc with a massive fragment might have a similar double peaked profile to the SED of a more evolved disc that contains a gap. However, in this case, it will be caused by the presence of an additional maximum at shorter wavelengths and will be similar only when the massive fragment is relatively cold ($\sim$400~K).
\end{abstract}

\begin{keywords}
{protoplanetary discs  -- planet-disc interactions -- planets and satellites: formation -- (planetary systems:)}
\end{keywords}

\maketitle%
\section{Introduction}

During the last decade, models for the formation and evolution of planetary systems have been developed significantly.
Early stages of planetary system formation have been studied by many authors (e.g. \citealt{Stamatellos09}; \citealt{Meru2010}; \citealt{Fouchet2010}; \citealt{Dodson11}; \citealt{Pinilla12,Pinilla15}; \citealt{Meru2014}; \citealt{Dipierro16}). Observational properties of the early stages of planetary and substellar companion formation and evolution are poorly understood. Authors mainly analyse the possibility to identify the signatures of a planet's formation with the best available techniques, such as ALMA -- Atacama Large Millimeter/submillimeter Array (\citealt{Gonzalez12}; \citealt{Vorobyov13}; \citealt{Zakhozhay13}; \citealt{Dipierro15}; \citealt{Dong16}) and IRAM Plateau de Bure interferometer (e.g. \citealt{Stamatellos11}). Recent ALMA observations of HL~Tau~\citep{ALMA15} and HW~Hya~\citep{Andrews16} very likely show that planetary formation is ongoing in the discs of these systems. The selection of a good target for observations is not a trivial task and it would be very useful to know the indicators that forming companions may leave in other more numerous observational methods, such as spectral energy distributions (SEDs), which are already available for many systems with protoplanetary discs. There are very few publications dedicated to this topic to date. \cite{Vorobyov13} and \cite{Zakhozhay13} show that a presence of hot and massive (32 -- 64~$M_J$) fragments -- proto-brown dwarfs during the first few 10,000 years -- initiate an additional peak in SEDs at 5 -- 10~$\mu$m. \cite{Varniere2006}, considered the example of a 2 Jupiter mass planet on a very close orbit to the star, and found that SEDs from discs with gaps may not only have a reduced emission at $\lambda\sim 5 - 20$~$\mu$m due to the removal of the emission from the gap, but also may have a measurable excess emission at longer wavelengths ($\sim 10 - 100$~$\mu$m), that rises from the heating of the vertical disc wall at the outer edge of the gap. \cite{Pinilla16} show that discs with massive planets lose the near infrared excess after a few Myrs and their SEDs look like true transition discs, while SEDs of the discs with less massive planets look like pre-transition disc SEDs for most of their remaining lifetimes.
\par There is a sufficient number of observational studies that contain large samples of observed SEDs of the systems with protoplanetary discs. For example, a recent paper of \cite{Marel16} contains a large sample of observational SEDs ($\sim$200 systems with transition discs), and $\sim$70$\%$ of these discs, most probably, contain large cavities and holes. These might be cleared by a planetary or a substellar companion or be a result of another physical process, like grain growth (e.g. \citealt{Dullemond05}) or photoevaporative clearing (e.g. \citealt{Alexander06}).

\par In this paper, we concentrate on the gaps cleared by one companion and analyse the possibility to detect signatures of the presence of a substellar or planetary companion in a protoplanetary disc based on SED profiles and also the possibility to determine the companion's parameters and the physical conditions of the disc when the presence of a companion is most evident.

\par This paper is organized as follows. The modeling approach and basic equations are briefly described in Section~\ref{algorithm}. The results and analysis are presented in Section~\ref{results}. Section~\ref{conclusions} contains the conclusions of the paper.

\section{Model assumptions}
\label{algorithm}

To calculate SEDs from the systems with protoplanetary discs and brown dwarf companions, we assume that the total flux from the system ($F$) is the sum of the fluxes from the central star ($F_{\ast}$), the disc ($F_d$) and the companion ($F_{c}$): $F = F_{d} + F_{c} + F_{\ast}$. Fluxes from the star and the companion are calculated using a black-body approximation:

\begin{equation}
F_{\ast} = \frac {\pi R_{\ast}^{2}}{d^2} B_{\nu} (T_{\ast}),
\end{equation}

\begin{equation}
F_{c} = \frac {\pi R_{c}^{2}}{d^2} B_{\nu} (T_{c,ef}),
\end{equation}
where $d$ is a distance to the object, $B_{\nu} (T)$ is the Plank function, $R_{\ast}$ and $T_{\ast}$ are the stellar radius and effective temperature, and $R_{c}$ and $T_{c, ef}$ are the companion's radius and effective temperature. The stellar and companion's physical parameters are taken from \cite{Baraffe15} and \cite{Baraffe98}, respectively, and are summarised in Table~\ref{tlab}.

\subsection{Spectral energy distribution from the disc}
\label{discSED}

\par To calculate the contribution to the SED from the protoplanetary disc, we assume that the companion is moving along a circular orbit and that there is no material inside the gap. The flux from the disc is $F_{d} = F_{disc} - F_{gap}$, where $F_{disc}$ is the flux from the disc, as it would be without the gap opened by the companion, and $F_{gap}$ is a flux from the part of the disc that has been cleared by the companion.
\par We use the solution to the one-dimensional radiative transfer equation to
calculate the protoplanetary disc flux density with the frequency $\nu$:
\begin{equation}
F_{disc} =d^{-2}\int_{R_{in}}^{R_{out}}B_{\nu}(T_r)~Q_{\nu}~2\pi r~{\rm d}r,
\end{equation}
\begin{equation}\label{form3}
Q_{\nu} = 1 - {\rm e}^{-\tau} ; ~~~ \tau = \varSigma_r~\kappa_{\nu},
\end{equation}
where $d$ is a distance to the system from the Sun, $r$ is the radial distance from the central object, $B_{\nu}(T_r)$ is the Planck function for a radius-dependent temperature, $T_r$, of the disc, $R_{in}$ and $R_{out}$ stand for the inner and outer radii of the disc, respectively, $Q_{\nu}$ represents the dust grain emission efficiency, $\tau$ stands for the optical depth of the disc material, which is the product of the wavelength-dependent disc opacity, $\kappa_{\nu}$, and $\varSigma_r$ is the radial surface density distribution of the disc material. To calculate the wavelength-dependent disc opacity, we use the Mie theory and consider compact spherical grains composed of astronomical silicates, a density of 2.5~g$\cdot$cm$^{-3}$, grain sizes between 0.1 and 100~$\mu$m and gas to dust mass ratio of 100.
\par Radial temperature and surface density distributions we assume to be power laws: $T_r \propto r^q$ and $\varSigma_r \propto r^p$. For computations we used $q$ = -0.5 and $p$ = -1.5. The effect of $q$ variation on the resulting SED profile is discussed in details in subsection~\ref{vs_disc}.
\par The disc surface density as a function of the disc mass is described by the equation:
\begin{equation}\label{Sigma_r}
\varSigma_r = \frac {M_d~r^p~(2+p)}{2 \pi~(R_{out}^{2+p} - R_{in}^{2+p})},
\end{equation}
where $M_d$ is a total mass of the disc.

\par The flux from the part of the disc that has been cleared by the companion, $F_{gap}$, is calculated using the same approach as $F_{disc}$, except only in the range $R_{in,gap}$ to $R_{out,gap}$, which are the gap inner and outer radii. These radii are determined by the distance to the star from the companion, $r_c$, and the Hill radius $R_H$:
\begin{equation}
R_H = r_c \Big (\frac{1}{3} \frac{M_c}{M_{\ast} + M_c} \Big)^{\frac{1}{3}},
\end{equation}
where $M_c$ and $M_{\ast}$ are masses of the companion and of the star, respectively.

\par We neglect the additional heating from the companion to the inner and outer edges of the disc gap. As discussed in detail in Appendix~\ref{Additional heat}, the maximum temperature to which the brown dwarf companion could heat the inner and outer sides of the disc gap is comparable, or much smaller, to the disc temperature to which it is been heated by the central star.

\section{Results and analysis}
\label{results}
We simulated synthetic SEDs for a system with a 0.8~$M_{\odot}$ central object and a 30~$M_J$ substellar companion with an age of 5~Myr. The mass of the companion we chose based on the limitation from the previous study of \cite{MaGe2014}, that shows that the maximum mass of the object that can form via  gravitational instability in a protostellar disc is $\sim40M_{J}$. Table~\ref{tlab} summarizes stellar and companion physical parameters that were used to compute the SED. The SED for the protoplanetary disc was modeled for $R_{in}$ = $R_{sub}$ (in our particular case $R_{sub}$ = 0.02~AU = 4.7~$R_{\ast}$) and $R_{out}$ = 400~AU. We consider a passive disc and assume there is no material inside the gap, and the companion is moving along circular orbit. We further assume that the width of the gap is one diameter of the Hill sphere, so we consider the minimum possible gap cleared by a companion of a given mass. 
\par We don't account for an increase of the emission temperature from the part of the outer edge of the gap that is directly exposed to stellar radiation since its contribution would be very small. For example, if a gap has been cleared by a companion at 1~AU, the total area that is not covered by the inner edge of the gap is $\sim0.07$~AU$^2$ at the distance 1.23~AU. However if the disc has a non-zero inclination, the addition flux from the outer edge of the gap would be more noticeable. In that case, the radiation from this part of the disc would be directly exposed to the observer and then even a small change of the temperature may cause a noticeable change of the flux. This will be studied in a subsequent paper with more detailed disc structure and radiative transfer simulations.

\begin{table*}
\centering
\caption{The physical parameters of the system at 5 Myr$^1$.}
\label{tlab}
\begin{tabular}{@{}cccccccccccccc@{}}
\hline\hline
 \multicolumn{4}{c}{Star} &&  \multicolumn{4}{c}{Companion} &&\\ 
\hline
$M_{\ast}$ &  $T_{\ast}$&  $L_{\ast}$&  $R_{\ast}$ && $M_{c}$& $T_{c}$&  $L_{c}$&  $R_{c}$ && $R_{sub}^2$ \\ 

$(M_{\odot})$ & $(K)$ & $(L_{\ast})$ & $(R_{\odot})$ && $(M_{J})$ & $(K)$ & $(L_{\ast})$ & $(\times 10^{9} cm)$ & & AU \\
\hline
  0.8 & 4081 & 0.426 & 1.306 && 30 & 2616 & $5.5 \times 10^{-3}$ & 25.27 && 0.02 \\
\hline\hline
\end{tabular}
\begin{minipage}{150 mm}
$^1$~Stellar parameters are from \citep{Baraffe15}, companion's parameters are from \citep{Baraffe98}
$^2$~$R_{sub}$ is determined using $T_{\ast}$ and $R_{\ast}$ and based on the assumption that disc dust sublimation temperature equals 1500~K, as in \cite{Dullemond01}.\\
\end{minipage}
\end{table*}

\par
\begin{figure}
\begin{center}
\includegraphics[width=\columnwidth]{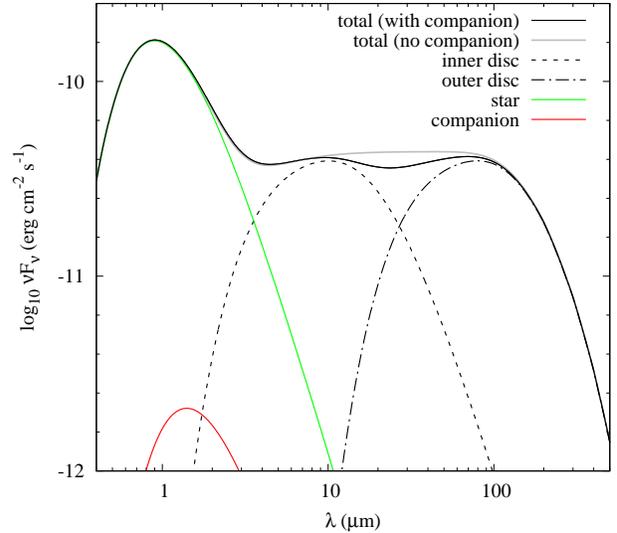}
\end{center}
\caption{
 SEDs of the modeled system with protoplanetary disc with an embedded companion (black solid line) that is composed of the inner (black dashed line) and outer (black dashed-dotted line) disc parts, the flux from the star (green line) and the flux of the companion (red line). The gray solid line shows the flux from the system with the same parameters but without a companion. We assume $d = 250$~pc.
}
\label{prim}
\end{figure}

 Figure~\ref{prim} shows SEDs from the system with a companion at 1~AU (black solid line) and without companion (gray solid line). The black dashed and dashed-dotted lines show the fluxes from the inner and outer parts of the disc. The red and green lines show the fluxes from the companion and the star, respectively.

\par Visual examination of Figure~\ref{prim} indicates that the presence of the gap cleared by the substellar companion along its orbital motion in the disc causes an additional minimum at the wavelength interval from 10 to 100~$\mu$m. Table~\ref{dif} contains the difference of the fluxes from the systems with and without the companion, and the absolute values for the fluxes for the each wavelength at the interval 10 -- 160~$\mu$m. The maximum difference $\sim70$~mJy corresponds to 33.8~$\mu$m. In the present study, we have analyzed the fluxes, assuming that the system is located at 250~pc. If the system is located closer to us, then the fluxes would be much more intense and hence the difference would be much more evident. For example, if the system is located at 100~pc, than all the fluxes in Table~\ref{dif} should be multiplied by the factor of 6.25.
\par In the following subsection, we analyse in more detail how the shape of SED profile and the difference between the fluxes with and without the companion depend on some of the companion and disc physical properties.

\begin{table}
\centering
\caption{The differences of the fluxes from the system with and without an embedded companion as a function of wavelength.}
\label{dif}
\begin{tabular}{@{}cccc@{}}
\hline\hline
$\lambda$  & $\Delta F$ & $F_{no\ companion}$ &  $F_{with\ companion}$  \\ 
  ($\mu$m)   &    (mJy)    &          (mJy)       &          (mJy)           \\
\hline
10 & 3.3 & 139.1 & 135.8 \\
20 & 46.2 & 288.4 & 242.2 \\
34 $(\Delta F_{max})^1$ & 70.0 & 490.7 & 420.7\\
80 & 36.2 & 1129.2 & 1093.0 \\
100 & 26.7 & 1316.3 & 1289.6 \\
160 & 12.8 & 1411.8 & 1399.0 \\
\hline\hline
\end{tabular}
$^1$ 34 $\mu$m is the wavelength where is the maximum difference 
between the fluxes of the system with and without companion.\\
\end{table}

\subsection{Dependence of SED profile from the properties of the protoplanetary discs and the companion.}
\label{protoplanetaryD}

\subsubsection{SED profile shape versus companion's location and mass}
\label{vs_comp}
\begin{figure}
\begin{center}
\includegraphics[width=\columnwidth]{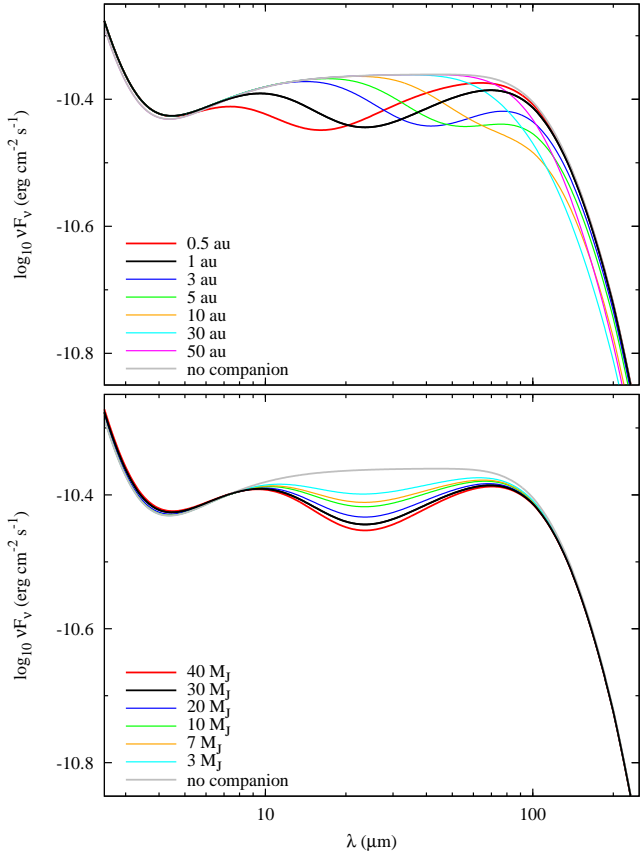}
\end{center}
\caption{
SEDs of the systems with companions at different distances from the star ({\it top panel}) and with companions of different masses (bottom panel). In top panel, SEDs of the model system with companion of $M_{c} = 30$~$M_J$ at different distances: 0.5~AU - red line, 1~AU - black line, 3~AU - blue line, 5~AU - green line, 10~AU - orange line, 30~AU - cyan line, 50~AU - magenta line. On bottom panel, SEDs of the modeled system with companion at 1~AU with different masses: 40~$M_J$ - red line, 30~$M_J$ - black line, 20~$M_J$ - blue line, 10~$M_J$ - green line, 7~$M_J$ - orange line, 3~$M_J$ - cyan line. On both panels, modeled SED from the system with the same parameters but without companion is shown with gray line. 
}
\label{rp_var}
\end{figure}

\begin{figure}
\begin{center}
\includegraphics[width=\columnwidth]{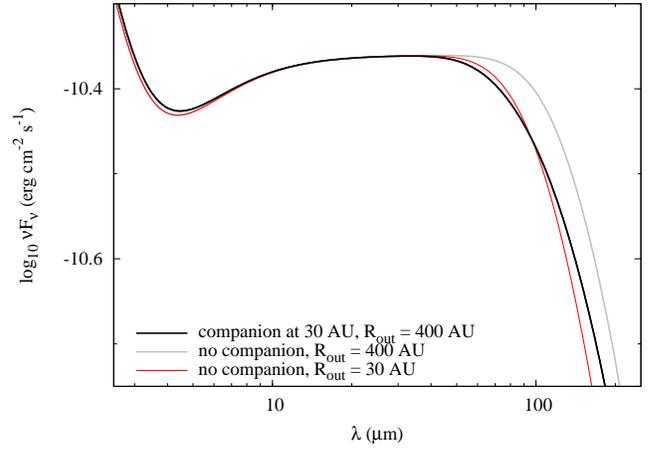}
\end{center}
\caption{
SEDs of the model system with protoplanetary disc with embedded brown dwarf at $r_c =$ 30~AU, $R_{out}$ = 400~AU (black line), SED of the analogous system but without companion (gray line) and from the system without companion and smaller $R_{out}$ = 30~AU (red line).
}
\label{Mp_short}
\end{figure}

Figure~\ref{prim} shows the SEDs from the system with an embedded companion in a protoplanetary disc located at 1~AU, because in this case the difference between the fluxes from the systems with and without the companion is the most evident. The distance of 1~AU we determined with the models by varying the distance from the companion to the central star within all possible positions along the disc radius (from 1~AU to 100~AU with steps of 1~AU). Figure~\ref{rp_var} illustrates the dependence of the shape of SED profile from the companion's location and mass, assuming all the parameters, except the one that we are varying, are the same as for the case shown in Figure~\ref{prim}. It is evident that the presence of the companion would be easier to determine if it would be located closer to the star. This is not because of the difference in fluxes, but because the presence of the companion at $<$10~AU causes a noticeable depression in the SED profile. The upper panel of Figure~\ref{rp_var} shows that that the biggest visual difference in intensities between SED profiles of the system with and without the embedded companion is for the system with the companion that is located very close to the inner edge of the disc. While computations show that for the systems with companions located closer to the star, the difference between the systems with and without the companion is smaller. For example, if the companion is located at 0.5~AU, the maximum difference is 49.4~mJy at 23.8~$\mu$m and if it is located at 10~AU, the maximum difference is 221.1~mJy at 106.9~$\mu$m. The largest difference between the fluxes with and without the companion is 266.7~mJy, for a companion at 30~AU from the star. But because this difference is at $\lambda = 154.5$~$\mu$m, where the disc is optically thin, visually it will be impossible to detect the signature of the companion because the SED profile resembling the SED of the system with the similar but shorter disc. Figure~\ref{Mp_short} shows the fluxes from the system with the disc and no companion, $R_{out}$ = 400~AU (gray line), with the same disc that contains a gap cleared by a companion ($M_c = 30$~$M_J$) at 30~AU (black line) and the flux from the shorter disc without companion, $R_{out}$ = 30~AU (red line). In the wavelength interval $\lambda \simeq 6 - 40$~$\mu$m fluxes from the system with the companion at 30~AU and system with shorter disc visually indistinguishable. At longer wavelengths, the flux from the system with a shorter disc is more intense at $\lambda < 100 \mu$m and less intense at $\lambda > 100$~$\mu$m, because the short disc has a significantly smaller amount of cold material comparing to the similar disc of typical size.

\par The lower panel of Figure~\ref{rp_var} illustrates the predictable dependence that more massive companions create deeper minima. The values of $M_c$ are varied from 40~$M_J$ down to 3~$M_J$. For all considered cases, companions are located at 1~AU and that is why the biggest difference between all the fluxes is at 33.8~$\mu$m. The maximum difference is for the system with the companion of 40~$M_J$ is - 76.7~mJy and a smallest difference of 32.8~mJy corresponds to the system with less massive companion of 3~$M_J$.

\par In both panels of figure~\ref{rp_var} at the wavelength $\lambda< 6$~$\mu$m, the spread of the lines is slightly noticeable. This difference is due to the contribution of the flux from the companion and it is non-negligible only for very massive companions with $M_c > 10$~$M_J$. For the companion of 30~$M_J$, it exceeds 1~mJy  in the wavelength interval $\lambda \thickapprox 1.43 - 2.76$~$\mu$m, and for the most massive companion considered in this work, 40~$M_J$, in the interval $\lambda \thickapprox 1.04 - 3.95$~$\mu$m. The maximum difference due to the flux from the brown dwarf companions with mass 30~$M_J$ and 40~$M_J$ are 1.14~mJy at $\lambda = 1.94$~$\mu$m and 1.70~mJy at $\lambda = 1.86$~$\mu$m, respectively.
Although, it is important to note that the additional flux from the companion is computed based on the assumption that the companion has the same age as the star and that there is no gas accretion. If the gas from the disc is still accreting onto the companion, the emission from the companion can be noticeably more intense. This will be studied in a future paper with more detailed structure of the disc, the embedded companion, and the gap.

\par The physical parameters for the companions with different masses from \cite{Baraffe98} and $R_H$, that were used to perform the computations are listed in Table~\ref{Mvar} of appendix~\ref{AdditionalSysPar}. 

\subsubsection{SED profile shape versus disc inner radius and temperature profile}
\label{vs_disc} 
\begin{figure}
\begin{center}
\includegraphics[width=\columnwidth]{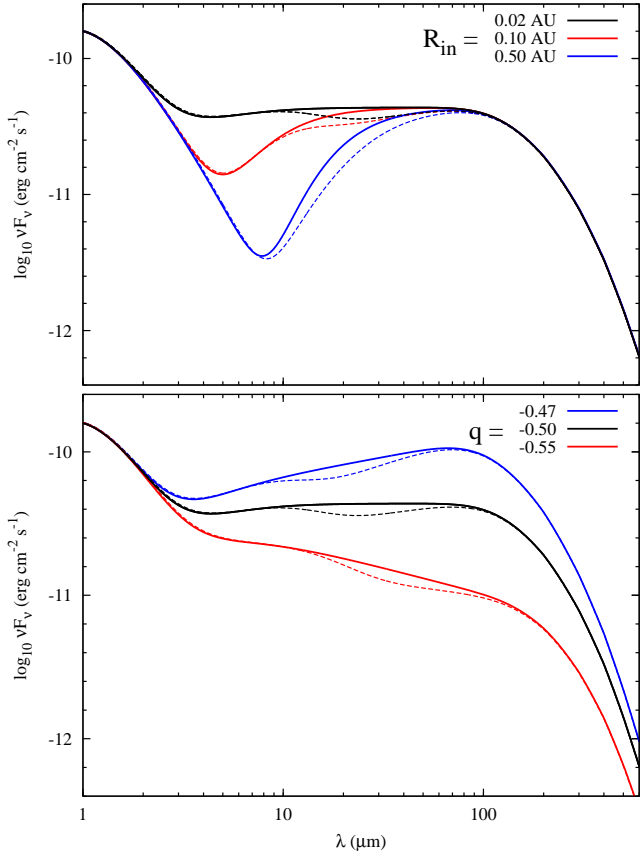}
\end{center}
\caption{
SEDs of the model system with an embedded brown dwarf at 1~AU with different $R_{in}$ ({\it top panel}): 0.02~AU (dust sublimation radius) -- black, 0.1~AU -- red and 0.5~AU -- blue dashed lines.
{\it Bottom panel} shows SEDs of the models with protoplanetary discs that have different temperature distribution indices: $q$ = -0.47 (blue), $q$ = -0.5 (black), $q$ = -0.55 (red). On both panels, SEDs from the system with a companion ($M_{c} = 30$~$M_J$, $r_c$ = 1~AU) are shown with dashed lines and solid lines show the fluxes from corresponding discs without companions.
}
\label{Rin_var}
\end{figure}

In \cite{Woitke16}, the authors showed the effect of different dust and disc parameters on the shape of model SED profile. In the present paper, we consider only two disc parameters: the location of the disc inner radius and a power law of the disc temperature profile. The top panel of Figure~\ref{Rin_var} shows SEDs from the systems with and without companions that have $R_{in}$ = 0.02, 0.1 and 0.5~AU (all another parameters are the same as in Figure~\ref{prim}). The value of $R_{in}$ = 0.02~AU corresponds to the dust sublimation radius \citep{Dullemond01}. The variation of $R_{in}$ doesn't change the difference of the fluxes for the systems with and without companions, or the wavelength of the maximum difference. Although the upper panel of Figure~\ref{Rin_var} indicates that if a companion's orbit is very close to the disc inner radius, like in the case of $R_{in}$ = 0.5~AU (blue lines), then the flux from the inner part of the disc is quite faint and the total SED from the system with a companion doesn't have a double peak profile. That is why the SED from such systems is very similar to the SED from the system with a slightly larger inner radius. On the other hand, for young discs with very small inner holes (with the values close to the sublimation radius), emission from the inner part of the disc (before the gap) will dominate at the wavelength interval $\lambda = 10 - 20$~$\mu$m. In the present paper, we consider quite dense discs that are optically thick at all the wavelengths up to the distances of few 10 of~AU, but if the inner disc would be optically thin then its emission at this wavelength interval would be determined by the chemical composition of the disc.
\par The lower panel of Figure~\ref{Rin_var} illustrates how the SEDs of the systems with and without companions depend on the temperature distribution profile power law $q$ ($T_r \propto r^q$). Obviously SED profiles strongly depend on the $q$ value. Although the difference between the fluxes from the systems with and without companions still corresponds to the wavelength interval $\lambda = 10 - 100$~$\mu$m. 
If the temperature profile is slightly shallower ($q$ = -0.47), then the disc would be hotter around the region of the gap and the maximum difference between the fluxes from the systems with and without companions will be larger, 110.6~mJy, and will correspond to a shorter wavelength $\lambda = 28.9$~$\mu$m. If the disc has a steeper temperature profile ($q$ = -0.55), then maximum difference from the fluxes would be 32.6~mJy at $\lambda = 43.3$~$\mu$m. Interestingly, the latter difference is comparable with the maximum difference from the system with a companion of 3~$M_J$ (32.8~mJy at $\lambda = 33.8$~$\mu$m), considered above and graphically presented in the lower panel of Figure~\ref{rp_var}. It additionally confirms the importance of the disc parameter determination, which would require a big observation data sample, corresponding to the wide wavelength interval.

\subsection{Comparison with another stages of disc evolution}
\label{parameters}

In this subsection, we compare the companion's formation signatures in SED profiles of the systems with protoplanetary and protostellar discs and compare our modelling results to those obtained before in \cite{Vorobyov13}. 
These models were based on numerical hydrodynamics simulations of \cite{Vorobyov10}, who studied the formation and evolution of protostellar discs subject to gravitational instability and fragmentation. The basic equations of hydrodynamics were solved on a polar grid in the thin-disc limit. That allowed to follow the gravitational collapse of a pre-stellar condensation (core) into the star plus disc formation stage and further to the T~Tauri stage when most of the parental core has accreted onto the burgeoning disc. The following physical processes were taken into account: disc self-gravity via solution of the Poisson integral and disc viscosity via $\alpha$-parameterization, radiative cooling from the disc suface, stellar and background irradiation, and also viscous and shock heating (for more details see \citealt{Vorobyov10}). In \cite{Vorobyov13}, the authors considered four moments in time: these were 0.09~Myr, 0.1~Myr, 1.1~Myr and 1.3~Myr since the formation of the central protostar. We choose an age of 0.1~Myr based on the similarity with the masses of the companions considered in the present paper. 

\par
\begin{figure}
\begin{center}
\includegraphics[width=\columnwidth]{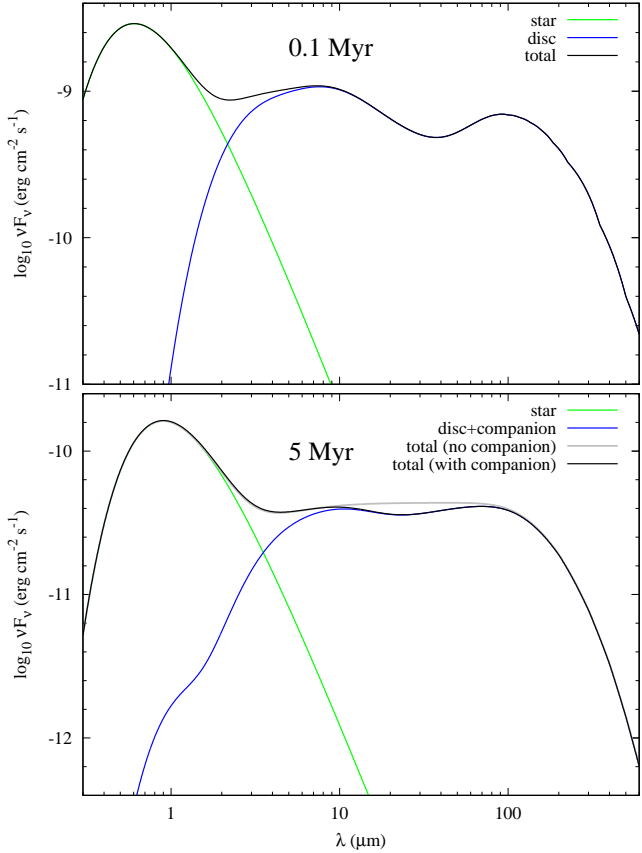}
\end{center}
\caption{SED of modeled systems with protostellar ({\it top panel} \citealt{Vorobyov13}) and protoplanetary ({\it bottom panel}) discs. In both panels, the flux from the star (protostar, at 0.1~Myr) is shown with green lines, the fluxes from the disc (including the flux from companion or a fragment, at 0.1~Myr) is shown with the blue line and the black line is a total flux from the system. On the bottom plot, flux from the corresponding system but without a companion is shown by the gray line. All discs have a face-on orientation and are located at 250~pc. The system's physical and geometrical parameters are described in the text and in Table~\ref{tlab}.
}
\label{age_var}
\end{figure}

 Figure~\ref{age_var} shows SEDs of a system with the protostar at 0.1~Myr (top panel, \citealt{Vorobyov13}) and with the star at 5~Myr (bottom panel) with surrounding discs and forming companions. The hot and most massive (32~$M_{J}$) fragment at 0.1~Myr is located at a radial distance of 108~AU. The embedded companion (30~$M_{J}$) at 5~Myr is located at 1~AU, because at this distance, the difference between the SED profiles of the system with and without the embedded companion is much more evident. In both panels, the total fluxes from the systems are shown with black lines, the fluxes from the star (5~Myr) or protostar (0.1~Myr) are shown with green lines, and the blue line shows the flux from the disc together with the companion. A slightly noticeable peak around 1~$\mu m$ from the disc at 5~Myr is due to the direct emission from the brown dwarf. This flux does not account for possible additional flux due to the gas accreation from the disc.

\par
\begin{figure}
\begin{center}
\includegraphics[width=\columnwidth]{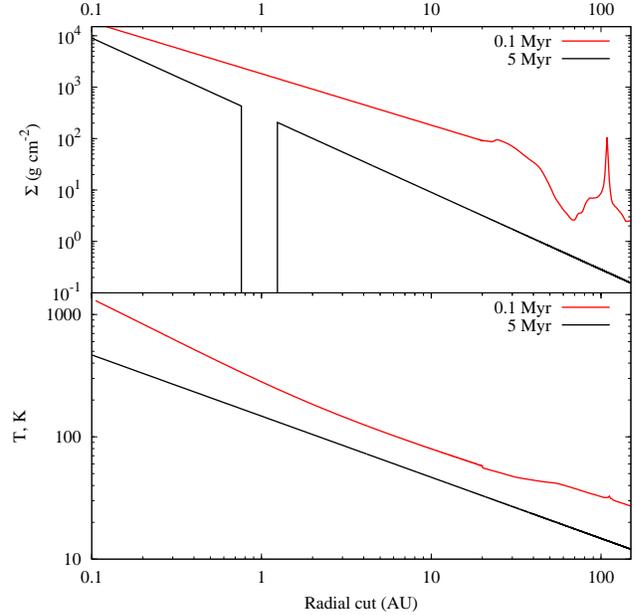}
\end{center}
\caption{
Azimuthally averaged surface density ({\it top panel}) and temperature ({\it bottom panel}). The red line shows surface density and temperature of the protostellar disc (as it was previously presented in \citealt{Vorobyov13}) and the black line is for the protoplanetary disc.
}
\label{Tsdens}
\end{figure}

 As studied in details in \cite{Vorobyov13}, the formation of a hot and massive fragment in the first 100k years causes an additional peak to be present at 5 -- 10~$\mu$m. The mid-plane temperature of the fragment at that moment in time, presented in Figure~\ref{age_var}, is 1180~K and the surface temperature is 380~K. Such mid-plane temperatures are too cold to form a companion, and that is why it is impossible to make a direct comparison between the different stages of the brown dwarf formation. An important conclusion that we can make by analysing Figure~\ref{age_var} is that SEDs of fragmenting discs and of the older disc with an embedded companion may have a similar double peak profile, caused by different physical processes: an additional peak due to the presence of a very hot clump in the protostellar disc and an addition minima due to the gap cleared by a companion in the protoplanetary disc. Figure~\ref{Tsdens} shows azimuthally averaged surface densities (top panel) and temperatures (bottom panel) as functions of the disc radial distance for the considered protostellar (red lines) and protoplanetary (black line) discs on both panels. The upper panel of Figure~\ref{Tsdens} illustrates the drop of surface density down to 0 from 0.77 to 0.23~AU for protoplanetary disc and an addition peak around 108~AU for protostellar discs - both these features are related to the double peak profiles of SEDs, that we see in figure~\ref{age_var}. 

\par The results obtained previously in \citep{Zakhozhay15} for a debris disc with a $40$~$M_{J}$ companion (the closest mass of the object, available in the \cite{Baraffe98} isochrones to the maximum mass of the object that could form in circumstellar disc \cite{MaGe2014}) indicate that the maximum difference between SEDs of the systems with and without a companion at $\approx 75$~$\mu$m is $\sim 10$~mJy (assuming $d$ = 250~pc) which is comparable with the sensitivity of the instrument PACS of the space telescope HERSCHEL in this wavelength regime. Although, in that paper, only one critical case was considered, assuming that the width for the gap is the diameter of one Hill sphere. This assumption is acceptable for the early stages of the disc evolution, like in a present paper, but for older debris discs, the gap cleared by the same companion should be wider. In debris discs, the width of the gap is determined by gravitational perturbations from the companion. Planetesimals entering the region around the companion's orbit (known as the "chaotic zone") are scattered onto highly eccentric orbits, creating an underdensity of the material (\citealt{Chirikov1979}; \citealt{Wisdom1980}). Recently, \citet{Nesvold2015} showed that collisions between planetesimals additionally widen the gap. In this case, the gap width would be at least 2-3 times wider than the one determined by the Hill radius. In a future investigation, we intend to extend our study, performing a similar analysis for debris discs with a wide range of different physical parameters.

\section{Conclusions}
\label{conclusions}

\par In this work, we present a study of SED profiles from a system with a protoplanetary disc that contains an embedded companion. We simulate the SEDs using a simplified flux computation approach and analyzed the effect of the most important disc and companion parameters, that affect the resulting profiles the most. We consider a system with a low mass star (0.8~$M_{\odot}$) and a brown dwarf companion. We find that the gap cleaned by the companion creates an additional depression in the SED profile at $\lambda \sim 10 - 100$~$\mu$m. The drop of the flux intensity strongly depends on the companion's mass and location:
1) a more massive companion initiates the deeper minimum;
2) the difference between the profiles is noticeable only if the companion is close enough (within $\sim$10~AU) to the star.
 
\par The analysis presented in subsection~\ref{protoplanetaryD} proves the importance of the disc parameters. It is shown that the maximum difference between the fluxes from the disc with and without a companion of 3~$M_J$ is similar to the difference between the systems with and without a companion of 30~$M_J$ but with a slightly steeper temperature profile ($q$ = -0.55). These maximum differences correspond to 33.4~$\mu$m and 43.3~$\mu$m, for 3~$M_J$ ($q$ = -0.50) and 30~$M_J$ ($q$ = -0.55), respectively, while the maximum flux difference is the same $\sim$33~mJy. Computations also indicate that if the disc inner radius is $>$0.5~AU, then the disc SED doesn't have a double peak profile. Thus the results of this paper are mainly related to the early stages of disc evolution when the companion just started to form a gap. On the transitional stage of the disc evolution the inner disc hole, as well as a gap width will be significantly larger and, hence, the evidence of the gap in the disc will be more evident in the SED profile (assuming that the embedded companion is at the same distance). Although at this stage it will be much harder to make a connection between the width of the gap and the mass of the companion, because at longer time intervals more physical processes may affect it.

\par We also made a comparison with SEDs from the protostellar disc with a hot fragment. Hot and massive fragments in a protostellar disc cause additional peaks in SED profile of the system. The intensity and wavelength regimes of the emission from the fragments depend on their temperature, density and spatial sizes. If the fragment is not very hot, like in \cite{Vorobyov13}, it will initiate an additional peak at 5 -- 10~$\mu$m that makes a total SED profile from the disc quite similar to what we have obtained in the present work for a protoplanetary disc with an embedded companion. Future modellings of SEDs from protostellar discs will allow us to determine the criteria that will help to distinguish these two cases. Although it is obvious that the most important is an age determination, because massive fragments have their very hot temperature and huge spacial sizes only at the very beginning of the disc formation. 

\par In the subsequent paper we intent to perform a more detailed analysis by using a more precise disc model and computing the disc temperature with the radiative transfer simulations. We will discuss in more detail how the disc chemical composition and vertical structure may affect the SED of the disc with a gap caused by presence of companions.

\begin{acknowledgements}
The author thanks the anonymous referee for useful suggestions that improved the manuscript. The author thanks F.~M\'{e}nard, I.~Kamp, A.~Carmona, D.~Asmus, P.~Pinilla, N.~van~der~Marel, C.~Dullemond, C.~Johnstone and P.~Berczik for useful discussions and Y.~Boehler for providing the dust opacities. Author also thanks the head of the research program $"$Fundamental properties of the material in the wide scale interval of the space and time Branch of Physics and Astronomy, National Academy of Sciences of Ukraine$"$, in the framework of which a part of this work was carried out.
\end{acknowledgements}

\begin{appendix}

\section{Additional heat of the gap rims from the brown dwarf companion}
\label{Additional heat}

\par
\begin{figure}
\begin{center}
\includegraphics[width=\columnwidth]{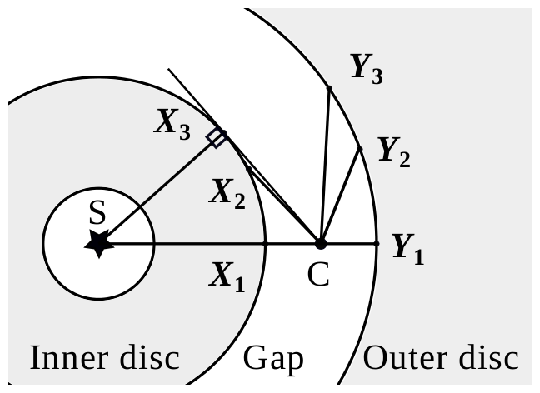}
\end{center}
\caption{
The schematic view of the star ($S$), the disc and the companion ($C$). $X_{1-3}$ and $Y_{1-3}$ are the points at the inner and outer edges of the rim, respectively (see the text for details).
}
\label{A1}
\end{figure}

 To estimate the additional heat that gets to the inner and outer edges of the gap of the disc from the substellar companion, let us consider the configuration presented in Figure~\ref{A1}. The temperature of every point of the gap edge is
\begin{equation}
T^4 = T^4_{rim} + T^4_{c,irr},
\end{equation}
where $T_{rim}$ is the temperature of the gap edge, which is heated by energy from the star that is absorbed and re-emitted by disc material, and $T_{c,irr}$ is the temperature to which the companion heats the edge of the gap, which is given by
\begin{equation}
T_{c,irr} = \sqrt [4] {\frac{L_c}{16 \pi \sigma r_{c,g}^2}},
\end{equation}
where $L_c$ is the luminosity of the companion and $r_{c,g}$ is the distance from the companion to the considered point of the gap edge. At the regions where the edge of the gap doesn't get the direct emission from the companion, $T = T_{rim}$. 
\par The minimum $r_{c,g}$ is equal to $R_H$, the point where the line connecting the star and the companion is perpendicular to the tangent to the circle which describes the inner and outer edges of the gap. In our particular case ($M_{\ast} = 0.8 M_{\odot}$, $M_c = 30$~$M_J$ and $r_c$ = 1~AU), it is 0.23~AU (at Figure~\ref{A1} points $X_1$ and $Y_1$ are shown for the inner and outer edges, respectively). In Figure~\ref{A1}, $S$ stands for the star and $C$ for the companion. Additionally, let us consider two more points of the inner and outer edges. 
\par $X_3$ is the furthermost point at the inner edge of the gap that is heated by the companion. This point is the intersection of the normal drawn from the point C to the circle describing the inner edge of the cavity with this circle. To determine the distance from the companion to this point, let us consider the right triangle $\bigtriangleup SX_3C$. The angle $SX_3C$ is right, $X_3S = r_c - R_H$, $CS = R_H$, and hence
\begin{equation}
CX_3 = \sqrt{2 r_c R_H - R_H^2}.
\end{equation}
In our particular case, $CX_3 =0.64$ AU. $X_2$ is the middle point and the distance to it from the companion is $X_2C = (X_3C+X_1C)/2 = 0.435$ AU. $Y_3$ and $Y_2$ are equidistant points to $X_3$ and $X_2$, respectively, at the outer edge of the gap.
\par In Table~\ref{A1}, we list the resulting temperatures ($T$) at points $X_1$, $X_2$, $X_3$, $Y_1$, $Y_2$, $Y_3$, together with the temperatures at the inner and outer edges without additional heat for the protoplanetary disc ($T_{rim}$), with physical parameters considered in the present paper and the irradiation temperature from the companion ($T_{c,irr}$). As one can see, $T_{c,irr}$ exceeds $T_{rim}$ only in one case at the point $Y_1$, where $r_{c,g}$ has a minimum value, and at further distances it is significantly smaller. The resulting temperatures ($T$) significantly exceed the temperatures of the rims only when the distance to the companion is the minimum: 15\%  and 30\% for the inner and outer rims, respectively. And for the further points, the increase in the temperatures due to the heat from the companion at $X_3$ and $Y_3$ is only 2\% and 6\%, respectively. Additional heat to the outer edge of the rim is bigger, because the outer edge is further away from the star and it is always colder. If the companion would be located further out from the star, the disc would be colder and the addition heat would be more noticeable, but the diameter of the companion's Hill sphere will increase as well and hence the resulting temperature increment will be still very small. 

\begin{table}
\centering
\caption{Temperatures at the gap rims that account for additional heat from the brown dwarf companion ($T$) and the corresponding temperatures of the rims ($T_{rim}$), irradiation temperatures from the companion ($T_{c,irr}$) and the distances between the companion and corresponding point ($r_{c,g}$).}
\label{rvar}
\begin{tabular}{@{}ccccc@{}}
\hline\hline
point  & $T$ & $T_{rim}$ & $T_{c,irr}$& $r_{c,g}$ \\
 & (K) & (K)&(K)  & (AU)\\  
\hline
  $X_1$ & 195 & 169 & 158 & 0.23 \\
  $X_2$ & 177 & 169 & 115 & 0.44 \\
  $X_3$ & 173 & 169 & 95  & 0.64 \\
  $Y_1$ & 174 & 132 & 158 & 0.23 \\
  $Y_2$ & 148 & 132 & 115 & 0.44 \\
  $Y_3$ & 140 & 132 & 95  & 0.64 \\

\hline\hline
\end{tabular}
\end{table}

\end{appendix}
\begin{appendix}
\section{Additional system parameters}
\label{AdditionalSysPar}

\par Table~\ref{Mvar} summarizes the physical parameters for the companions with different masses \citep{Baraffe98} with corresponding Hill radii for the $r_c$=~1~AU, that were used to compute SEDs presented in the bottom panel of Figure~\ref{rp_var}.

\begin{table}
\centering
\caption{The physical parameters of the companions with different masses \citep{Baraffe98} and corresponding Hill radii (for $r_c$ = 1~AU),
that were used for Figure~\ref{rp_var}}
\label{Mvar}
\begin{tabular}{@{}cccc@{}}
\hline\hline
 $M_c$  & $T_{c}$ & $R_{c}$ &  $R_H$  \\ 
$(M_J)$ &   (K)    & $(\times 10^{9} cm)$ & (AU) \\
\hline
3 & 1098 & 10.27 & 0.108\\
7 & 1668 & 11.54 & 0.142\\
10 & 1965 & 12.62 & 0.160\\
20 & 2452 & 18.18 & 0.201\\
30 & 2616 &25.27& 0.229\\
40 & 2754 & 28.52 & 0.251\\
\hline\hline
\end{tabular}
\end{table}

\end{appendix}


\begin{thebibliography}{}

\bibitem[\protect\citeauthoryear{ALMA Partnership}{2015}]{ALMA15}
ALMA Partnership et al., 2015, ApJL, 808, 1, L3

\bibitem[\protect\citeauthoryear{Alexander et al.}{2006}]{Alexander06}
Alexander R.D., Clarke C.J., \& Pringle J.E., 2006, MNRAS, 369, 216

\bibitem[\protect\citeauthoryear{Andrews et al.}{2016}]{Andrews16}
Andrews S.M. et al., 2016, ApJL, accepted

\bibitem[\protect\citeauthoryear{Baraffe et al.}{1998}]{Baraffe98} 
Baraffe I., Chabrier G., Allard F., Hauschildt P.H., 1998, A\&A, 337, 403

\bibitem[\protect\citeauthoryear{Baraffe et al.}{2015}]{Baraffe15} 
Baraffe I., Homeier D., Allard F., Chabrier G., 2015, A\&A, 577, A42

\bibitem[\protect\citeauthoryear{Chirikov}{1979}]{Chirikov1979}
Chirikov B. V., 1979, PhR, 52, 263

\bibitem[\protect\citeauthoryear{Dipierro et al.}{2015}]{Dipierro15}
Dipierro G., Price D., Laibe G., Hirsh K., Cerioli A., Lodato G., 2015, MNRAS, 453, L73

\bibitem[\protect\citeauthoryear{Dipierro et al.}{2016}]{Dipierro16}
Dipierro G., Laibe G., Price D.J., Lodato G., 2016, MNRAS, 459, L1

\bibitem[\protect\citeauthoryear{Dodson-Robinson \& Salyk}{2011}]{Dodson11}
Dodson-Robinson S.E. \& Salyk C., 2011, ApJ, 738, 131

\bibitem[\protect\citeauthoryear{Dong et al.}{2016}]{Dong16}
Dong R., Vorobyov E., Pavlyuchenkov Y., Chiang E., Liu H.B., 2016, ApJ, 823, 141

\bibitem[\protect\citeauthoryear{Dullemond \& Dominik}{2005}]{Dullemond05}
Dullemond C. P. \& Dominik C., 2005, A\&A, 434, 971

\bibitem[\protect\citeauthoryear{Dullemond et al$.$}{2001}]{Dullemond01} 
Dullemond C.P., Dominik C., Natta A., 2001, ApJ, 560, 957

\bibitem[\protect\citeauthoryear{Fouchet et al.}{2010}]{Fouchet2010}
Fouchet L., Gonzalez J.-F., \& Maddison S.T., 2010, A\&A, 518, A16

\bibitem[\protect\citeauthoryear{Gonzalez et al.}{2012}]{Gonzalez12}
Gonzalez J.-F., Pinte C., Maddison S.T., M\'{e}nard F., \& Fouchet L.,  2012, A\&A, 547, A58

\bibitem[\protect\citeauthoryear{Ma \& Ge}{2014}]{MaGe2014} 
Ma B., Ge J., 2014, MNRAS, 439, 2781

\bibitem[\protect\citeauthoryear{Meru \& Bate}{2010}]{Meru2010}
Meru F. \& Bate M.R., 2010, MNRAS, 406, 2279

\bibitem[\protect\citeauthoryear{Meru et al.}{2014}]{Meru2014}
Meru F., Quanz S.P., Reggiani M., Baruteau C., Pineda J.E. arXiv:1411.5366

\bibitem[\protect\citeauthoryear{Nesvold \& Kuchner}{2015}]{Nesvold2015}
Nesvold E.R. \& Kuchner M.J., 2015, ApJ, 798, 83

\bibitem[\protect\citeauthoryear{Pinilla et al.}{2012}]{Pinilla12}
Pinilla P., Benisty M., and Birnstiel T., 2012, A\&A, 545, A81 

\bibitem[\protect\citeauthoryear{Pinilla et al.}{2015}]{Pinilla15}
Pinilla P., de Juan Ovelar M., Ataiee S., Benisty M., Birnstiel T., van Dishoeck E. F., Min M., 2015, A\&A, 573, A9

\bibitem[\protect\citeauthoryear{Pinilla et al.}{2016}]{Pinilla16}
Pinilla P., Klarmann L., Birnstiel T., Benisty M., Dominik C., Dullemond C.P., 2016, A\&A, 585, A35

\bibitem[\protect\citeauthoryear{Stamatellos \& Whitworth}{2009}]{Stamatellos09}
Stamatellos D. \& Whitworth A.P., 2009, MNRAS, 392, 413

\bibitem[\protect\citeauthoryear{Stamatellos et al.}{2011}]{Stamatellos11}
Stamatellos D., Maury A., Whitworth A., Andr\`{e} P., 2011, MNRAS, 413, 1787

\bibitem[\protect\citeauthoryear{van der Marel et al.}{2016}]{Marel16}
van der Marel N., Verhaar B.W., van Terwisga S., Merin B., Herczeg G., Ligterink N.F.W., van Dishoeck E.F., 2016, A\&A, accepted

\bibitem[\protect\citeauthoryear{Varni\`{e}re et al.}{2006}]{Varniere2006}
Varni\`{e}re P., Bjorkman J.E., Frank A. et al., 2006, ApJ, 637, L125  

\bibitem[\protect\citeauthoryear{Vorobyov et al.}{2010}]{Vorobyov10} 
Vorobyov E.I., Basu S., 2010, ApJ, 719, 1896

\bibitem[\protect\citeauthoryear{Vorobyov et al.}{2013}]{Vorobyov13} 
Vorobyov E.I., Zakhozhay O.V., \& Dunham M.M., 2013, MNRAS, 433, 3256

\bibitem[\protect\citeauthoryear{Wisdom}{1980}]{Wisdom1980}
Wisdom J., 1980, AJ, 85, 1122

\bibitem[\protect\citeauthoryear{Woitke et al.}{2016}]{Woitke16}
Woitke et al., 2016, A\&A, 586, A103

\bibitem[\protect\citeauthoryear{Zakhozhay et al.}{2013}]{Zakhozhay13}
Zakhozhay O.V., Vorobyov E.I., \& Dunham M.M., 2013, Memorie della Societ\`{a} Astronomica Italiana, 84, 4, 880

\bibitem[\protect\citeauthoryear{Zakhozhay}{2015}]{Zakhozhay15}
Zakhozhay O.V., 2015, Kinematics and Physics of Celestial Bodies, 31, 4, 184



\end{thebibliography}
\end{document}